\documentclass[aps,prd,twocolumn,nofootinbib,longbibliography,10pt]{revtex4}
\usepackage{etoolbox}
\usepackage{dcolumn,tensor,nicefrac}
\usepackage{amsmath,amssymb,amsfonts,mathtools}
\usepackage{mathrsfs,bbold}
\usepackage{graphicx}
\usepackage{subcaption}
\usepackage[colorlinks=true,urlcolor=blue,citecolor=red,linkcolor=blue]{hyperref}
\usepackage{accents}
\newlength{\dhatheight}
\usepackage{natbib}
\usepackage{wrapfig}
\usepackage{booktabs}
\usepackage{float}
\usepackage{flushend,BOONDOX-cal,BOONDOX-frak}

\makeatletter
\newsavebox{\@brx}
\newcommand{\llangle}[1][]{\savebox{\@brx}{\(\m@th{#1\langle}\)}%
  \mathopen{\copy\@brx\kern-0.5\wd\@brx\usebox{\@brx}}}
\newcommand{\rrangle}[1][]{\savebox{\@brx}{\(\m@th{#1\rangle}\)}%
  \mathclose{\copy\@brx\kern-0.5\wd\@brx\usebox{\@brx}}}
\makeatother
\begin{document}
\title{\textbf{Quasi-normal modes of quantum gravity black hole with perfectly fluid dark matter}}
\author{Arpita Jana}
\email{janaarpita2001@gmail.com}
\affiliation{Department of Astrophysics and High Energy Physics, S. N. Bose National Centre for Basic Sciences, JD Block, Sector-III, Salt Lake City, Kolkata-700 106, India}
\author{Manjari Dutta}
\email{chandromouli15@gmail.com}
\affiliation{Department of Astrophysics and High Energy Physics, S. N. Bose National Centre for Basic Sciences, JD Block, Sector-III, Salt Lake City, Kolkata-700 106, India}
\author{Sunandan Gangopadhyay}
\email{sunandan.gangopadhyay@gmail.com}
\affiliation{Department of Astrophysics and High Energy Physics, S. N. Bose National Centre for Basic Sciences, JD Block, Sector-III, Salt Lake City, Kolkata-700 106, India}
\begin{abstract}
\noindent In this work, we have studied the motion of a massless scalar photon in the renormalization group (RG)
improved Schwarzschild black hole spacetime in the presence of perfectly fluid dark matter (PFDM).
Considering the critical orbit conditions and the null geodesics condition in static spherically sym-
In metric geometry, we have shown the variation of the radius of the photon sphere $r_{ph}$ with the PFDM
parameter $\zeta$. Due to perturbations in black hole spacetime, gravitational waves are emitted in the
form of quasi-normal radiations, which correspond to quasi-normal modes (QNMs). In this work,
we have studied two types of perturbations in RG improved Schwarzschild spacetime: scalar field
perturbations and electromagnetic (EM) field perturbations. For both cases, we have studied
the effect of the PFDM parameter on the quasi-normal mode frequencies and the shadow of the black
hole, which is related to the photon radius.
\end{abstract}
\maketitle
\section{introduction}
\noindent Black holes are the most fascinating and mysterious predictions of Einstein's theory of general relativity\cite{Einstein1,Einstein2}. 
The first experimental confirmation of the prediction of GR was the observation of light deflection during the solar eclipse in 1919. Actually, the gravitational force of a black hole is so intense that any object moving around it within a certain radius falls into it. These effects have a combined name called \textit{gravitational lensing} \cite{schneider1,blandford,petters,vperlick,schneider2,bartelmann,dodelson,congdon,perlick}.
After a century of first observations of light deflection due to gravity, the existence of the black holes has also been verified in some recent observations \cite{abbott,akiyama1,akiyama2,akiyama3} by the LIGO and VIRGO observatories in collaboration with the Event Horizon Telescope (EHT).
\noindent They published the image of the shadow of a supermassive M87 black hole \cite{akiyama1,akiyama2,akiyama3}.\\
\noindent If light passes close to a black hole, the rays can either be deflected or can travel on a circular orbit. Thus, the light coming from a source behind the black hole can cast a shadow of the black hole on a certain plane, which is visible to a distant observer. Theoretical analyses of black hole shadow have been started in \cite{synge} for spherically symmetric black holes. Later in \cite{luminet}, the formation of shadows for spherically symmetric black holes surrounded by accretion disks has been studied. The analysis for the shadow of rotating black holes has been discussed in \cite{chandrashekhar,falcke}. In \cite{falcke,meliafalcke,bardeen,hiokimaeda,vries}, the first idea of observing a black hole shadow was produced by numerical simulations. Since then, with the improvements in observational data, various studies on black hole shadows have been done in modified theories of gravity and wormholes as well as higher dimensional gravity theories \cite{grezenbach1,grezenbach2,ovgun1,ovgun2,weizou,abdujabbarov1,wangchen,mishrachak,kumarsingh1,kumarsingh2,hennigar,khodabakhshi,amirghosh,shaikh,contreras,lulyu,fenglu,kumarghosh1,kumarghosh2,malu,konoplya1,dastan1,dastan2,dsg1,smg,amarilla1,amarilla2,amarilla3,vetsov,wangxu,zhuwu,paponi,abdujabbarov2,singhghosh,perlicktsupko,liguo}.\\
\noindent The study of black hole spacetime and the motion of any particle moving in its vicinity has impactful insights on the features of black holes, such as the Hawking temperature. Particles moving in the black hole spacetime follow the geodesics, in the absence of external forces. The geodesics around any black hole can form closed or open orbits which may be stable or unstable depending on the potential barrier experienced by the particles.  The stability of the orbits can be understood by using the Lyapunov exponents \cite{lyapunov,pikovsky}. These exponents can be expressed in terms of the second order derivative of the effective potential barrier at extrema \cite{pradhan}.\\
\noindent The evolution of a binary black hole system proceeds through three distinct phases: inspiral, merger, and ringdown. The inspiral phase provides information about the mass and spin, using post-Newtonian approximation \cite{blanchet}. The merger phase describes the collapse of two objects leading to the formation of black holes \cite{pretorius,campanelli,baker}. The final phase describes the perturbed black hole emitting gravitational waves (GW) in terms of quasi-normal radiation \cite{berti}.  This perturbations influence the black hole spacetime and its stability. The frequencies of these perturbations are named as \textit{quasi-normal frequencies (QNF)} and corresponding modes are known as \textit{quasi-normal modes (QNM)}. Several analytical and numerical studies of QNMs have been done in \cite{ferrari,schutzwill,iyerwill,konoplya2,konoplya3,konoplya4,chandrashekhar2,leaver,horowitz1,chocornell,kokkotas,berti2,toshmatov,toshmatov2,momennia,ding,churilova1,churilova2,bsalcedo,bronnikov,eniceicu,caimiao,panotop,turimovzhidenko}.\\
\noindent In this paper, we have considered two types of field perturbations in the RG improved Schwarzschild black hole spacetime. First one is the scalar field perturbations and the second one is the electromagnetic field perturbations. We have studied the effects of these fields on black hole spacetime individually in the presence of PFDM. Due to these perturbations, the forms of effective potential barrier change accordingly for both scalar and EM fields. Using these forms, we have studied the effect of PFDM on the QNFs which are related to the effective potential calculated at $r_{ph}$. Considering the null geodesics and circular orbit condition, we have shown the variation of the size of the photon sphere $r_{ph}$ with PFDM parameter $\zeta$.  Along with these, we have presented a comparison study between the quasi-normal frequencies of two types of field perturbations in the above mentioned black hole geometry in the presence of PFDM. \\
\noindent The arrangement of the paper is as follows. In section II, we have presented a basic review of the geodesics of a massless scalar photon in a static, spherically symmetric black hole spacetime, leading to the equation for the photon radius. In section III, we have discussed the theoretical construction of black hole shadow and its relation with the photon sphere. In section IV, we have presented the derivation of the metric of Schwarzschild black hole in the presence of PFDM and then we have provided the brief description of renormalization group improvement of the metric. In section V, we have revisited the derivation of quasi-normal frequencies for spherically symmetric black hole geometries and presented the effect of PFDM parameter on the effective potential for both scalar and electromagnetic perturbations. In section VI, we have presented the effect of PFDM on QNFs and black hole shadow, by plotting the parameters against the PFDM parameter ($\zeta$). In the final section, we have discussed the important findings of this work.    

\section{trajectory of a photon in spherically symmetric static spacetime}
\noindent The metric of a static spherically symmetric black hole in (3+1)-dimension can be written as
\begin{equation}\label{metric}
    ds^{2}=-f(r)dt^{2}+f(r)^{-1}dr^{2}+r^{2}\left(d\theta^{2}+\sin^{2}{\theta d\phi^{2}} \right)
\end{equation}
where $f(r)$ denotes the lapse function of the black hole. To study about the geodesics of a massless photon in this geometry, we shall start with the Lagrangian of the system which has the form
\begin{equation}
    \mathcal{L}=\frac{1}{2}g_{\mu\nu}\dot{x}^{\mu}\dot{x}^{\nu}
\end{equation}
where the partial derivatives are defined with respect to the affine parameter $\lambda$.
Trajectories of any massless particle in any spacetime geometry correspond to the null geodesics which satisfy the geodesic condition 
\begin{equation}\label{nullgeodesic}
    g_{\mu\nu}\dot{x}^{\mu}\dot{x}^{\nu}=0~.
\end{equation}
In order to obtain the general trajectory of photon in the black hole spacetime, we need to use the separability approach of the Hamilton-Jacobi equation mentioned in \cite{Carter,paponi}. The Hamilton-Jacobi equation with spacetime metric $g_{\mu \nu}$ has the general form as \cite{Carter}
\begin{equation}\label{jacobieq}
    \frac{\partial S}{\partial \lambda}=-\frac{1}{2}g^{\mu \nu}\frac{\partial S}{\partial x^{\mu}} \frac{\partial S}{\partial x^{\nu}}
\end{equation}
where $\lambda$ is the affine parameter. The Hamilton-Jacobi action $S$ can be written as
\begin{equation}\label{action}
    S=\frac{1}{2}m^{2}\lambda -Et+L\phi +S_{r}(r)+S_{\theta}(\theta)
\end{equation}
to separate the equations of motion for different coordinates.\\
\noindent For photon ($m=0$), using eq.(\ref{jacobieq}) and eq.(\ref{action}), we obtain
\begin{equation}\label{carter}
\frac{r^{2}E^{2}}{f(r)}-r^{2}f(r)\left( \frac{\partial S_{r}}{\partial r}\right)^{2}=\left( \frac{\partial S_{\theta}}{\partial \theta}\right)^{2}+\frac{L^{2}}{\sin^{2}{\theta}}~.
\end{equation}
Here, we have used the metric components from eq.(\ref{metric}); $g^{tt}=-\frac{1}{f(r)}, g^{rr}=f(r), g^{\theta \theta}=\frac{1}{r^{2}}$ and $ g^{\phi \phi}=\frac{1}{r^{2}\sin^{2}{\theta}}$.\\ \noindent Introducing the separation constant $\mathcal{K}$, we obtain from eq.(\ref{carter})
\begin{align}
    & \left( \frac{\partial S_{r}}{\partial r}\right)^{2}=\frac{E^{2}}{(f(r))^{2}}-\frac{\mathcal{K}^{2}}{r^{2}f(r)},\nonumber \\ & 
    \left( \frac{\partial S_{\theta}}{\partial \theta}\right)^{2}=\mathcal{K}-\frac{L^{2}}{\sin^{2}{\theta}}~.
\end{align}
We have considered the static and spherically symmetric geometry. Since the metric is independent of $t$ and $\phi$ coordinates, the generalized momenta corresponding to these are conserved. 
Now to obtain the trajectory of the photon in the black hole spacetime, we shall use the generalized momentum, $p_{\mu}= \frac{\partial S}{\partial x^{\mu}}=g_{\mu\nu}\frac{dx^{\nu}}{d\lambda}$. Hence the equations of motion of the photon obtained as
\begin{align}\label{eom}
   & \frac{dt}{d\lambda}=\frac{E}{f(r)}, ~~~~~~
   \frac{dr}{d\lambda}=\frac{1}{r^{2}}\sqrt{\left(E^{2}r^{4}-\mathcal{K}r^{2}f(r) \right)},\nonumber \\& \frac{d\theta}{d\lambda}= \frac{1}{r^{2}}\sqrt{\left(\mathcal{K}- \frac{L^{2}}{\sin^{2}{\theta}} \right)},~~~ \frac{d\phi}{d\lambda}=\frac{L}{r^{2}\sin^{2}{\theta}}~.
\end{align}
The radial geodesics can be expressed as 
\begin{equation}
    \left(\frac{dr}{d\lambda}\right)^{2}-V(r)=0
\end{equation}
where $V(r)=E^{2}\left(1-\frac{\mathcal{K}f(r)}{r^{2}E^{2}} \right)$ and it is called the effective radial potential. We need to find the critical orbit for the photon from the unstable condition
\begin{equation}\label{criticalorbit}
    V(r=r_{ph})=0, ~~~~~~\frac{\partial V}{\partial r}\biggr|_{r=r_{ph}}=0,~~~~~ \frac{\partial^{2} V}{\partial r^{2}}\biggr|_{r=r_{ph}}< 0
\end{equation}
where $r_{ph}$ is the radius of the critical orbit which is called the photon sphere. We shall study the null geodesics here. Hence using the conditions mention in eq.(\ref{criticalorbit}), we can obtain 
\begin{equation}\label{photonradius}
    \frac{2}{r_{ph}}f(r_{ph})-f^{\prime}(r_{ph})=0~.
\end{equation}
Here $f^{\prime}(r_{ph})$ denotes the derivative of $f(r)$ with respect to $r$ at $r=r_{ph}$. Solving eq.(\ref{photonradius}) for any black hole geometry, we can get the expression for the radius of the photon sphere.\\
\noindent Now using eq.(\ref{nullgeodesic}) and eq.(\ref{photonradius}), we can obtain the ratio of angular momentum and the energy of the photon as
\begin{equation}
    \frac{L}{E}=\frac{r_{ph}}{\sqrt{f(r_{ph})}}~.
\end{equation}

\section{Black hole shadow: construction and relation with photon sphere}
\noindent In this section, we shall discuss the construction of black hole shadow and how it is related with the photon sphere. The detailed explanation about the construction of black hole shadow is described in \cite{perlick}.
\begin{figure}[H]
\begin{center}
\includegraphics[scale=0.31]{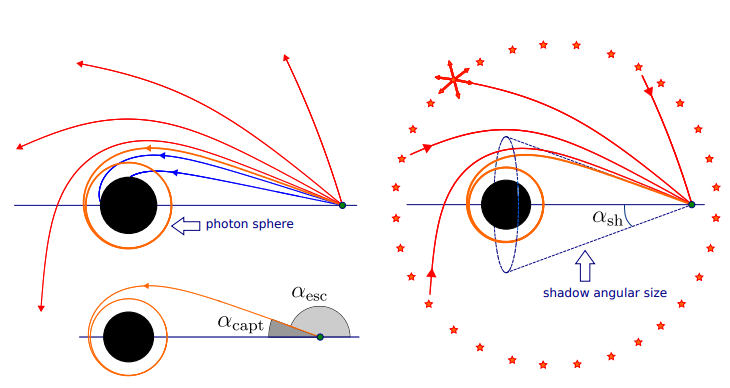}
\caption{Construction of black hole shadow \cite{perlick}\label{diag2}.}
\end{center}
\end{figure}
\noindent The light rays approaching a black hole from a distant source can follow two distinct trajectories : some are scattered and escape to infinity after being deflected by the gravitational field of the black hole; while others are trapped and fall through the event horizon of the black hole. The critical boundary separating these two behaviors defines the photon sphere where light orbits the black hole under the intense gravitational force.   Now, consider an idealized scenario in which light sources are uniformly distributed throughout the universe, except in the region between the observer and the black hole. In this case, only the deflected light rays can reach the observer, while those captured by the event horizon cannot. Consequently, the region corresponding to the captured rays appears dark to the observer. This dark area constitutes the shadow of the black hole.
\begin{figure}[H]
\begin{center}
\includegraphics[scale=0.4]{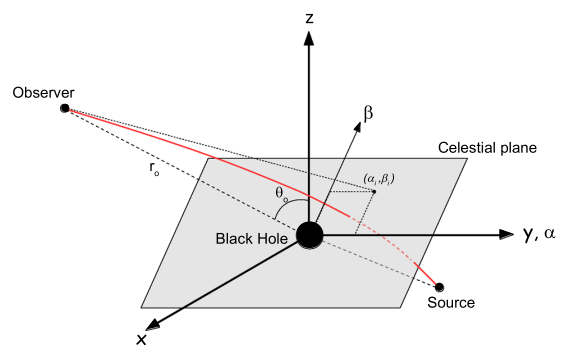}
\caption{Diagram for celestial coordinates $\alpha$ and $\beta$ \cite{smg}\label{diag1}.}
\end{center}
\end{figure}

\noindent To obtain the size of the black hole shadow, we have to write down the celestial coordinates $\alpha$ and $\beta$ \cite{vazquez}. In (3+1)-dimension, for a distant observer, the celestial coordinates are defined as \cite{dassahagangopadhyay}
\begin{align}\label{celestialdef}
    & \alpha = \lim_{r \rightarrow \infty} -r^{2}\sin{\theta} \frac{d\phi}{dr} \biggr|_{r_{ph},\theta=\pi/2}, \nonumber \\ &
    \beta = \lim_{r \rightarrow \infty} r^{2}\frac{d\theta}{dr} \biggr|_{r_{ph},\theta=\pi/2}~.
\end{align}
Here $r$ is the distance between the black hole and the observer, $\alpha$ is the apparent perpendicular distance of the shadow from the symmetry axis and $\beta$ denotes the apparent perpendicular distance of the shadow from its projection on the equatorial plane. For the reference, see the Fig.(\ref{diag1}) taken from \cite{smg}.\\ 
Using the equations of motion mentioned in eq.(\ref{eom}), we obtain 
\begin{align}\label{thetaphi}
    & \frac{d\phi}{dr}=\frac{L}{\sin^{2}{\theta}\sqrt{E^{2}r^{4}-\mathcal{K}r^{2}f(r)}}~~~,\nonumber \\ &
    \frac{d\theta}{dr}=\sqrt{\frac{\mathcal{K}- \frac{L^{2}}{\sin^{2}{\theta}}}{E^{2}r^{4}-\mathcal{K}r^{2}f(r)}}~~~.
\end{align}
Now using eq.(\ref{thetaphi}) and eq.(\ref{celestialdef}) and taking the limit $r \rightarrow \infty$, we obtain the celestial coordinates as
\begin{equation}\label{celestial}
    \alpha=-\frac{L}{E}, ~ \beta=\sqrt{\frac{\mathcal{K}}{E^{2}}}~.
\end{equation}
The radius of the black hole shadow in $\alpha$-$\beta$ plane is 
\begin{equation}\label{shadowdef}
    R_{s}=\sqrt{\alpha^{2}+\beta^{2}}=\sqrt{\xi^{2}+\eta}
\end{equation}
where $\xi=\frac{L}{E}$ and $\eta=\frac{\mathcal{K}}{E^{2}}$. Using the critical orbit condition eq.(\ref{criticalorbit}), it is obtained that 
\begin{equation}\label{xietasum}
    \xi^{2}+\eta = \frac{r_{ph}^{2}}{f(r_{ph})}~.
\end{equation}
Now combining eq.(\ref{shadowdef}) and eq.(\ref{xietasum}), we obtain the shadow radius as
\begin{equation}\label{shadowradius}
    R_{s}=\frac{r_{ph}}{\sqrt{f(r_{ph})}}~.
\end{equation}
This equation defines the direct relationship between the photon radius $r_{ph}$ and the shadow radius of a black hole $R_{s}$.

\section{stability analysis of photon orbit : lyapunov exponent}
\noindent In this section, we shall discuss the brief review of the Lyapunov exponent. \\
\noindent Lyapunov exponents measure the average rate of convergence or divergence of nearby trajectories in phase-space. A positive Lyapunov exponent denotes a chaotic system and a negative Lyapunov exponent denotes a stable system. To study the geodesic analysis in terms of Lyapunov exponents, we shall start with equation of motion of a dynamical system given by \cite{cardosomiranda}
\begin{equation}
    \frac{dY_{i}}{dt}=G_{i}(Y_{j})~.
\end{equation}
Perturbing the system slightly, we can have (upto linear order)
\begin{equation}\label{linearizedeq}
    \frac{d(\delta Y_{i}(t))}{dt}=K_{ij}(t)\delta Y_{j}(t)
\end{equation}
where $K_{ij}$ is called the \textit{linear stability matrix}\cite{cornishlevin}, defined as
\begin{equation}
    K_{ij}(t)= \frac{\partial G_{i}}{\partial Y_{j}}\biggr |_{Y_{i}(t)}~.
\end{equation}
The solution of eq.(\ref{linearizedeq}) is written as
\begin{equation}
    \delta Y_{i}(t)=L_{ij}(t)\delta Y_{j}(t=0)
\end{equation}
with $L_{ij}$ being the \textit{evolution matrix} \cite{cardosomiranda} which obeys $L_{ij}(t=0)=\delta_{ij}$. \\
\noindent The eigenvalues of the linear stability matrix $K_{ij}$ give the Lyapunov exponents $\lambda$ which determines the stability of the system. \\
\noindent We are interested in the circular orbit of a massless photon in an equatorial plane. The analysis will be carried out in the phase-space ($p_{r}$, $r$) for coordinate time $t$. The Euler-Lagrange equation for $r$ is given by
\begin{equation}
    \frac{d}{d\lambda}\left(\frac{\partial \mathcal{L}}{\partial \dot{r}}\right)-\frac{\partial \mathcal{L}}{\partial r}=0~.
\end{equation}
Here $\dot{r}=\frac{dr}{d\lambda}$\footnote{All overdots denote the derivative with respect to the affine parameter $\lambda$.} and the generalized momentum is given by
\begin{equation}
    p_{r}=\frac{\partial \mathcal{L}}{\partial \dot{r}}~.
\end{equation}
Using the above equations, we obtain
\begin{align}
    & \frac{dr}{dt}=\frac{\dot{r}}{\dot{t}} \nonumber \\ &
    \frac{dp_{r}}{dt}=\frac{\dot{p}_{r}}{\dot{t}}~.
\end{align}
Now the system is perturbed with $r\rightarrow r+\delta r$ and $p_{r}\rightarrow p_{r}+\delta p_{r}$ and we obtain 
\begin{equation}
    \delta p_{r}=\left( \frac{\partial^{2}\mathcal{L}}{\partial \dot{r}^{2}} \right ) \delta \dot{r}
\end{equation}
and since $\dot{t}$ is fixed on circular orbit, we get
\begin{align}
   & \delta \left( \frac{dr}{dt}\right)= \frac{1}{\dot{t}} \left( \frac{\partial^{2}\mathcal{L}}{\partial \dot{r}^{2}} \right )^{-1} \delta p_{r} \nonumber \\ &
   \delta \left( \frac{dp_{r}}{dt} \right)=\frac{1}{\dot{t}} \frac{d}{dr}\left(\frac{\partial \mathcal{L}}{\partial r} \right) \delta r
   ~.
\end{align}
We can write the above set of equations in matrix form as
\begin{equation}
    \frac{d}{dt} 
    \begin{pmatrix}
        \delta p_{r}\\ \\
        \delta r
    \end{pmatrix} = \begin{bmatrix}
        0 & \frac{1}{\dot{t}} \frac{d}{dr}\left(\frac{\partial \mathcal{L}}{\partial r} \right) \\
        \frac{1}{\dot{t}} \left( \frac{\partial^{2}\mathcal{L}}{\partial \dot{r}^{2}} \right )^{-1} & 0
    \end{bmatrix}
    \begin{pmatrix}
        \delta p_{r}\\ \\
        \delta r
    \end{pmatrix}~.
\end{equation}
This matrix is the linear stability matrix and to find the Lyapunov exponents we, need to find the eigenvalues of the matrix evaluated on the photon sphere $r_{ph}$. Hence the Lyapunov exponent becomes
\begin{equation}
    \lambda^{2}=\frac{1}{g_{rr}\dot{t}^{2}}\frac{d}{dr}\left(\frac{\partial \mathcal{L}}{\partial r} \right) \biggr |_{r=r_{ph}}=\frac{\left(\frac{d^{2}V}{dr^{2}}\right)}{2\dot{t}^{2}}\biggr |_{r=r_{ph}}~.
\end{equation}
Here, we have used $\dot{r}^{2}=V(r)$ and the critical orbit condition mentioned in eq.(\ref{criticalorbit}).  

\section{renormalization group improvement of Schwarzschild black hole immersed in PFDM}
\noindent This section is based on the derivation of the metric for Schwarzschild black hole in the presence of PFDM and the RG improvement of this metric. 
\subsection{Schwarzschild black hole in the presence of PFDM}
\noindent Here we consider the (3+1)-dimensional gravity theory in the presence of PFDM \cite{kiselev1,kiselev2}. The action can be written as
\begin{equation}\label{action}
    I=\int d^{4}x \sqrt{-g}\left(\frac{R}{16\pi G}+\mathcal{L}_{PFDM} \right)
\end{equation}
where $\mathcal{L}$ is the Lagrangian density due to the presence of PFDM,$G$ is the Newton's gravitational constant and R is the Ricci scalar. Extremizing the action, we can get the corresponding equations of motion as
\begin{equation}\label{einsteineom}
    R_{\mu\nu}-\frac{1}{2}g_{\mu\nu}R=-8\pi G T_{\mu\nu}^{PFDM}
\end{equation}
with $R_{\mu\nu}$ being the Ricci scalar and $T_{\mu\nu}^{PFDM}$ being the energy-momentum tensor for the PFDM medium. The energy-momentum tensor for PFDM has the form \cite{xuwang,dasrc,dassahagangopadhyay}
\begin{equation}\label{emtensor}
    (T^{\mu}~_{\nu})^{PFDM}=diag(-\rho, P_{r},P,P)~~; P_{r}=-\rho
\end{equation}
with $\rho$ and $P$ being the energy density and the pressure of the medium respectively. In order to find the equation of state of the PFDM, we shall use the approach mentioned in \cite{liyang,xuwang} and hence we consider \cite{dassahagangopadhyay}
\begin{align}\label{assumption}
    &(T^{\theta}~_{\theta})^{PFDM}=(T^{t}~_{t})^{PFDM}(1-\epsilon)\nonumber \\ &(T^{\phi}~_{\phi})^{PFDM}=(T^{r}~_{r})^{PFDM}(1-\epsilon)
\end{align}
where $\epsilon$ is a constant. Substituting eq.(\ref{emtensor}) into eq.(\ref{assumption}), the equation of state for PFDM is obtained as \cite{xuwang}
\begin{equation}
    \frac{P}{\rho}=(\epsilon - 1)~.
\end{equation}
There can be many choices for the value of $\epsilon$ \cite{liyang} but in our case, we have chosen $\epsilon=\frac{3}{2}$ \cite{liyang,houxu}. Hence the equation of state of perfectly fluid dark matter becomes 
\begin{equation}\label{eos}
    \frac{P}{\rho}=\frac{1}{2}~.
\end{equation}
To solve the Einstein field equations, we shall take an ansatz for a static, spherically symmetric metric \cite{dassahagangopadhyay}
\begin{equation}\label{ansatzmetric}
    ds^{2}=-e^{\sigma(r)}dt^{2}+e^{\gamma(r)}dr^{2}+r^{2}(d\theta^{2}+\sin^{2}{\theta}d\phi^{2})~.
\end{equation}
Using eq.(\ref{einsteineom}) and eq.(\ref{ansatzmetric}), we obtain the equations as
\cite{dassahagangopadhyay}
\begin{align}\label{eoms}
    & e^{-\gamma}(\frac{1}{r^{2}}-\frac{\gamma^{\prime}}{r})-\frac{1}{r^{2}}=-8\pi G \rho,\nonumber \\ &
    e^{-\gamma}(\frac{1}{r^{2}}+\frac{\sigma^{\prime}}{r})-\frac{1}{r^{2}}=-8\pi G \rho,\nonumber \\ &
    \frac{e^{-\gamma}}{2}(\sigma^{\prime \prime}+\frac{\sigma^{\prime^{2}}}{2}+\frac{\sigma^{\prime}-\gamma^{\prime}}{r}-\frac{\sigma^{\prime}\gamma^{\prime}}{2})=8\pi G P, \nonumber \\&
    \frac{e^{-\gamma}}{2}(\sigma^{\prime \prime}+\frac{\sigma^{\prime^{2}}}{2}+\frac{\sigma^{\prime}-\gamma^{\prime}}{r}-\frac{\sigma^{\prime}\gamma^{\prime}}{2})=8\pi G P~.
\end{align}
In the above equations, prime and double prime denotes first and second derive with respect to $r$. Now, taking the ratio of the first and the third equations and using eq.(\ref{eos}), we obtain
\begin{equation}\label{ratio}
    \frac{e^{-\gamma}}{2}(\sigma^{\prime \prime}+\frac{\sigma^{\prime^{2}}}{2}+\frac{\sigma^{\prime}-\gamma^{\prime}}{r}-\frac{\sigma^{\prime}\gamma^{\prime}}{2})=-\frac{1}{2}\lbrace e^{-\gamma}(\frac{1}{r^{2}}-\frac{\gamma^{\prime}}{r})-\frac{1}{r^{2}} \rbrace~.
\end{equation}
Now subtracting the first two equations in eq.(\ref{eoms}), we obtain 
\begin{equation}
     \sigma^{\prime}+\gamma^{\prime}=0 \Rightarrow \sigma+\gamma=0
\end{equation}
where the integration constant has been set to zero.\\
\noindent We now define $\sigma(r)=\ln{(1-B(r))}$ and hence substituting it in eq.(\ref{ratio}), we get
\begin{equation}
    r^{2}B^{\prime \prime}+3rB^{\prime}+B = 0~.
\end{equation}
Solving this equation, we obtain
\begin{equation}
    B(r)=\frac{r_{s}}{r}-\frac{\zeta}{r}\ln{\left(\frac{r}{|\zeta|}\right)}
\end{equation}
with $r_{s}$ and $\zeta$ being the integration constant. Setting $\zeta=0$ and using weak field approximation, we obtain $r_{s}=2GM$. \\
\noindent Comparing eq.(\ref{metric}) and eq.(\ref{ansatzmetric}), the lapse function of the black hole is obtained as
\begin{equation}\label{lapse}
    f(r)=e^{\sigma(r)}=(1-B(r))=1-\frac{2GM}{r}+\frac{\zeta}{r}\ln{\left(\frac{r}{|\zeta|}\right)}~.
\end{equation}
Here $M$ is the mass of the black hole and $\zeta$ represents the intensity of the PFDM medium \cite{dassahagangopadhyay}.

\subsection{Metric for RG improved Schwarzschild black hole in the presence of PFDM}
\noindent As it is expected that the theory of general relativity would break down at very small length scale due to its nonrenormalizability \cite{desser,klauder}, the developing theory of quantum gravity is one of the fascinating topics to the researchers. One of the schemes is the asymptotic safety scenario which is based on the renormalization group (RG) approach \cite{reuter1,reuter2,percacci}. In 2000, Bonanno and Reuter \cite{bonannoreuter}, first applied this approach in a Schwarzschild black hole background, where they replaced the usual Newton's gravitational constant with the running gravitational constant. These type of renormalization group improved black holes are named as \textit{quantum corrected black holes}. The methods to achieve such quantum improved geometry have been discussed elaborately in \cite{bonannoreuter, PawlowskiStock, Platania, ReuterWeyer}. The first step is to obtain the momentum scale-dependent coupling constants from the renormalization group flow equations; while the next step is to write down the momentum cut-off scale in terms of the radial coordinates of spherically symmetric geometry. There are mostly used two ways to do the scale identification, one of which is based on constructing curvature scalars like Ricci scalar and Kretschmann scalar, while the other method is based on the UV fixed point which separates the weak-coupling and strong-coupling regime. The third step to achieve the geometry has to done in three possible ways. Firstly, one can directly impose the flowing coupling constants into the classical solution to improve that. Secondly, one can put the running coupling constants in the equation of motion; such as we can get the quantum improved Einstein field equation by imposing running gravitational constant and running cosmological constant
\begin{equation}
    G_{\mu \nu} = 8\pi G(x) T_{\mu \nu}-\Lambda(x)g_{\mu \nu}
\end{equation}
where the flow of the matter coupling has been ignored. The final approach is to impose the running couplings in the action level. The quantum improved Lagrangian density for Einstein-Hilbert action can be written as
\begin{equation}
    \mathcal{L}=\frac{\sqrt{-g}}{16\pi G(x)}(R-2\Lambda(x))~.
\end{equation}
The simplest way to achieve the geometry is to impose running couplings into the solution level\footnote{In this paper, all the analyses have been executed for $\Lambda=0$.}. This was a quick revisit of quantum corrected geometries.\\
\noindent For a quantum corrected Schwarzschild black hole, the lapse function can be written as
\begin{equation}\label{lapse1}
    f(r)=1-\frac{2G(r)M}{r}
\end{equation}
where $G(r)$ is running gravitational constant. The flow of Newton's gravitational constant $G$ has been originated from the renormalization group flow equation \cite{bonannoreuter,HarstReuter,EichhornVersteegen}
\begin{equation}{\label{beta}}
    k\frac{d\tilde{G}(k)}{dk} = 2\tilde{G}(k)\left( 1-\frac{\tilde{G}(k)}{4\pi \tilde{\alpha}}\right)
\end{equation}
with $\tilde{G}(k)$ being the dimensionless Newton's constant, defined as $\tilde{G}(k)=k^{2}G(k)$ and $\tilde{\alpha}$ is a constant which corresponds to the fixed point $\tilde{G}_{\star}=4\pi \tilde{\alpha}$. Integrating eq.(\ref{beta}) within the limit $0$ and $k$, we obtain the flow of Newton's gravitational constant in terms of momentum cut-off scale as \cite{Ishibashi}
\begin{equation}\label{kflow}
    G(k)=\frac{4\pi \tilde{\alpha}G}{4\pi \tilde{\alpha}+k^{2}G}~.
\end{equation}
Using the method of cut-off identification in terms of radial coordinate \cite{HarstReuter,EichhornVersteegen}, the position dependent flow of Newton's gravitational constant reads 
\begin{equation}\label{flow}
    G(r)= \frac{G}{1+\frac{\tilde{\omega}G}{r^{2}}}~.
\end{equation}
$\tilde{\omega}$ is the quantum gravity parameter. \\
\noindent Now the lapse function eq.(\ref{lapse1}) takes the form
\begin{equation}\label{lapsefunction}
    f(r)=1-\frac{2GMr}{(r^{2}+\tilde{\omega}G)}~.
\end{equation}
Hence, the metric for quantum corrected Schwarzschild black hole in the presence of PFDM can be written down using eq.(\ref{lapse}) and eq.(\ref{lapsefunction}) as
\begin{equation}\label{finalmetric}
    f(r)=1-\frac{2GMr}{(r^{2}+\tilde{\omega}G)}+\frac{\zeta}{r}\ln{\left(\frac{r}{|\zeta|}\right)}~.
\end{equation}
This is the metric which we shall use in our further analysis of this work.

\section{Quasi-normal modes of a field for quantum corrected schwarzschild black hole surrounded by pfdm}
\noindent In this section, we shall discuss about the perturbation of the static, spherically symmetric black hole geometry in the presence of a scalar field and electromagnetic field individually.
\subsection{Scalar perturbations}
\begin{center}
\begin{figure}[H]
    \centering
    \begin{subfigure}{0.45\textwidth}
        \centering
        \includegraphics[scale=0.47]{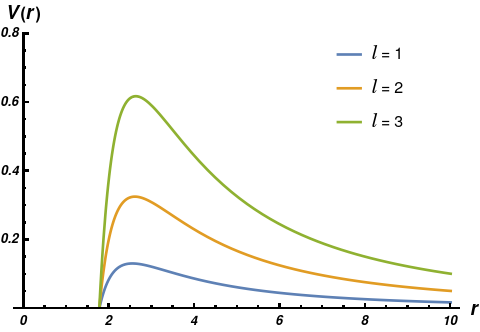}
        \caption{Effective potential for $l=1,2,3$ ($\zeta = 0.07$)}
        \label{plot1}
    \end{subfigure}
    \hfill
    \begin{subfigure}{0.45\textwidth}
        \centering
        \includegraphics[scale=0.47]{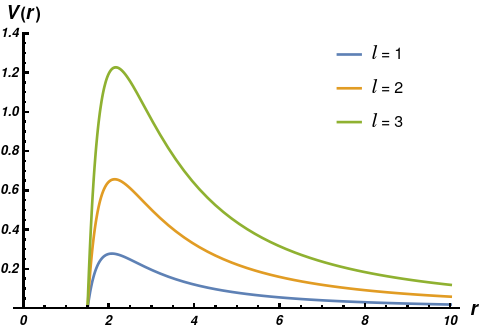}
        \caption{Effective potential for $l=1,2,3$ ($\zeta = 0.80$)}
        \label{plot2}
    \end{subfigure}
    \caption{Comparison of effective potentials for scalar field perturbations for $l = 1, 2, 3$ at different values of $\zeta$.}
    \label{combined_potential}
\end{figure}
\end{center}

\noindent The massless scalar field $\Phi$ in curved spacetime follows the Klein-Gordon equation
\begin{equation}\label{kg}
    \frac{1}{\sqrt{-g}}\partial_{\mu}\left(\sqrt{-g}~g^{\mu\nu}\partial_{\nu}\Phi \right)=0~.
\end{equation}
Here $\sqrt{-g}=r^{2}\sin{\theta}$ which is found from the determinant of the (3+1)-dimensional spacetime metric mentioned in eq.(\ref{metric}). The ansatz for the solution is given in terms of the spherical harmonics and has the from \cite{jusufi,fernando,arfken}
\begin{equation}
    \Phi(t,r,\theta,\phi)= e^{-i\omega t}Y_{l,m}(\theta,\phi)\frac{\psi(r)}{r}
\end{equation}
with $l=0,1,2,...$ being called as multipole number. After imposing separation of variables, the scalar field perturbation equation can be recast as
\begin{equation}\label{perturbeq}
    \frac{f(r)}{r^{2}}\frac{d}{dr}\left[r^{2}f(r)\frac{d}{dr}\left(\frac{\psi(r)}{r}\right) \right]+\left(\omega^{2}-f(r)\frac{l(l+1)}{r^{2}} \right)\frac{\psi(r)}{r}=0~.
\end{equation}
\noindent Now defining tortoise coordinate $r_{*}$ by the definition $\frac{dr_{*}}{dr}=\frac{1}{f(r)}$, eq.(\ref{perturbeq}) takes the form of Schr$\ddot{o}$dinger-like wave equation with potential $V(r_{*})$
\begin{equation}\label{waveeq}
    \frac{d^{2}\psi}{dr_{*}^{2}}+\left(\omega^{2}-V(r_{*}) \right)\psi=0
\end{equation}
where $V(r_{*})$ has the following form
\begin{equation}\label{potential}
    V(r_{*})=f(r)\left(\frac{l(l+1)}{r^{2}}+\frac{f^{\prime}(r)}{r} \right)~.
\end{equation}
\noindent Substituting the form of the lapse function $f(r)$ from eq.(\ref{finalmetric}), we get the effective potential as
\begin{widetext}
\begin{align}
    V(r) &= \left(1-\frac{2GMr}{(r^{2}+\tilde{\omega}G)}+\frac{\zeta}{r}\ln{\left(\frac{r}{|\zeta|}\right)} \right)\times \left(\frac{l(l+1)}{r^{2}}+\frac{2GM(r^{2}-\tilde{\omega}G)}{r(r^{2}+\tilde{\omega}G)^{2}}+\frac{\zeta}{r^{3}}\left(1-\ln{\left(\frac{r}{|\zeta|}\right)}\right) \right)\nonumber \\ &
    \approx \frac{l(l+1)}{r^{2}}+\frac{1}{r^{3}}\biggr[(l(l+1)-1)\left(\zeta \ln{\left( \frac{r}{|\zeta|}\right)}-2GM-1 \right) \biggr]+\frac{1}{r^{4}}\biggr[4GM\zeta \left( \ln{\left( \frac{r}{|\zeta|}\right)}-\frac{1}{2}\right)-\zeta^{2}\ln{\left( \frac{r}{|\zeta|}\right)}\left( \ln{\left( \frac{r}{|\zeta|}\right)}-1\right) \biggr]\nonumber \\ &
    ~~~~~~ + \frac{2}{r^{5}}G^{2}M\tilde{\omega}(l(l+1)-3)+\frac{1}{r^{6}}\biggr[16G^{3}M^{2}\tilde{\omega}-8G^{2}M\tilde{\omega}\zeta \left( \ln{\left( \frac{r}{|\zeta|}\right)}-\frac{1}{4}\right) \biggr]~.
\end{align}
\end{widetext}
\noindent In Fig.(\ref{combined_potential}), we have plotted the effective potential $V(r)$ versus $r$  for scalar field perturbations. To plot this, we have set $\tilde{\omega}=0.01$. It is clear from the plot that increasing the value of $l$ increases the height of the potential. Two plots are for two different values of PFDM parameter $\zeta$ and for large value of $\zeta$, potential has higher peak.
\onecolumngrid
\twocolumngrid
\noindent Since, eq.(\ref{waveeq}) is similar to the one-dimensional Schr$\ddot{o}$dinger equation, we shall use the WKB approximation method to solve this and this can be expressed in the form of
\begin{equation}\label{qmeq}
    \frac{d^{2}\psi(x)}{dx^{2}}+Q(x)\psi(x)=0~.
\end{equation}
In the black hole perturbation equation, $\psi$ represents the radial part of the wavefunction. In eq.(\ref{qmeq}), $x$ denotes the tortoise coordinate $r_{*}$ which ranges from $-\infty$ to $+\infty$ and $Q(x)\equiv \left(\omega^{2}-V(r_{*}) \right)$, which is constant at $x\rightarrow \pm \infty$. 
\begin{figure}[H]
\begin{center}
\includegraphics[scale=0.25]{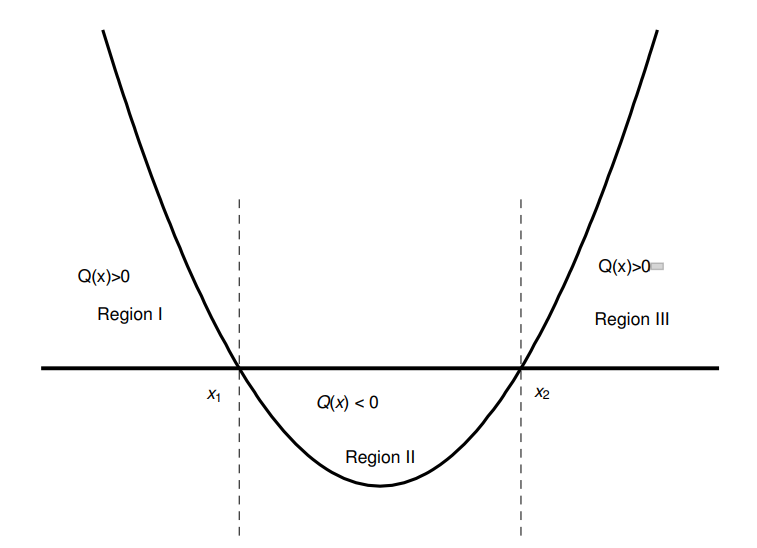}
\caption{behaviour of Q(x) \cite{dasrc,zettiliqm}\label{fig2}.}
\end{center}
\end{figure}
\noindent $Q(x)$ follows the behaviour shown in Fig.(\ref{fig2}).  The semi-analytical WKB approximation has been used in \cite{schutzwill} to solve this problem.\\
\noindent The WKB approximation method is used to solve a system with slowly varying potential; which means the potential remains almost constant over a region of the order of the de-Broglie wavelength. To solve eq.(\ref{qmeq}), we consider a trial form for $\psi(x)$ in terms of amplitude $A(x)$ and a phase factor $\theta(x)$\footnote{This $\theta$ and the spherical coordinate $\theta$ mentioned in eq. (\ref{metric}) are not same.}\cite{griffithqm,zettiliqm}
\begin{equation}\label{trialform}
    \psi(x)=A(x)e^{i\theta(x)}~.
\end{equation}
The amplitude $A(x)$ changes slowly with $x$. Using eq. (\ref{qmeq}) and (\ref{trialform}), we obtain the solution for $\psi(x)$ in the region where $Q(x)> 0 $ as \cite{schutzwill,griffithqm,zettiliqm}
\begin{align}
    & \psi(x)=\frac{1}{(Q(x))^{\frac{1}{4}}}~~\exp{\left(\pm ~ i \int \limits_{x}^{x_{1}}\sqrt{Q(x^{\prime})}~~ dx^{\prime} \right)}~; ~~(Region ~~I) \\ &
    \psi(x)=\frac{1}{(Q(x))^{\frac{1}{4}}}~~\exp{\left(\pm ~i \int \limits_{x_{2}}^{x}\sqrt{Q(x^{\prime})}~~ dx^{\prime} \right)}~; ~~(Region ~~III)
\end{align}
Assuming that the two turning points $x_{1}$ and $x_{2}$ are closely aligned that means $x_{1}\simeq x_{2}=x_{0}$, $Q(x)$ can be approximated as a parabola in region II where $Q(x)<0$. Hence, $Q(x)=Q_{0}+\frac{1}{2}(x-x_{0})^{2}\frac{d^{2}Q}{dx^{2}}|_{x=x_{0}}+...$ where $Q_{0}\equiv Q(x_{0})<0, Q^{\prime}(x_{0})=0$ and $Q^{\prime \prime}(x_{0})>0$. \\
\noindent Now eq.(\ref{qmeq}) can be written in the following form
\begin{equation}\label{diffeq}
    \frac{d^{2}\psi}{dx^{2}}+\left(Q_{0}+\frac{1}{2}Q_{0}^{\prime \prime}(x-x_{0})^{2} \right)\psi(x)=0~.
\end{equation}
Solving the above equation, we shall be able to obtain the form of $\psi(x)$ in the region where $Q(x)<0$. To find the exact solution, we shall use the parabolic cylinder functions \cite{bender}. Now defining
\begin{align}\label{functiondef}
    \kappa \equiv \frac{1}{2}Q_{0}^{\prime \prime},& ~~ q \equiv (4\kappa)^{1/4}e^{i\pi /4}(x-x_{0}),\nonumber \\& \mu+\frac{1}{2}\equiv \frac{-iQ_{0}}{\sqrt{2Q_{0}^{\prime \prime}}}
\end{align}
we can rewrite eq. (\ref{diffeq}) in the following form
\begin{equation}\label{paraboliceq}
    \frac{d^{2}\psi}{dq^{2}}+\left(\mu+\frac{1}{2}-\frac{q^{2}}{4} \right)\psi(q)=0~. 
\end{equation}
Solutions of this equation are parabolic cylinder functions $D_{\mu}(q)$. Now, following the method used in \cite{bender} we obtain the solution as
\begin{equation}
    \psi(q)\equiv D_{\mu}(q) = C_{1}q^{\mu}e^{-q^{2}/4}+C_{2}q^{-\mu-1}e^{q^{2}/4}~.
\end{equation}
Since only asymptotic solutions are expected, hence, $e^{q^{2}/4}$ term will blow up with $C_{2}$ being zero and $C_{1}=1$. Now to know the complete asymptotic expansion of $D_{\mu}(q)$, we consider
\begin{equation}\label{parabolicansatz}
    D_{\mu}(q) = q^{\mu}e^{-q^{2}/4}w(q)
\end{equation}
with $w(q)$ having an asymptotic power series in the order of $\frac{1}{q}$. Substituting eq.(\ref{parabolicansatz}) into eq.(\ref{paraboliceq}) and using the method of power series, the asymptotic expansion ($q\rightarrow\infty$) of $w(q)$ reads \cite{bender}
\begin{equation}\label{wexpansion}
    w(q)=1-\frac{\mu(\mu-1)}{2q^{2}}+\frac{\mu(\mu-1)(\mu-2)(\mu-3)}{8q^{4}}+...
\end{equation}
Now $D_{\mu}(q)$ can be written in terms of the linear combinations of the other two solutions of eq.(\ref{paraboliceq})
\begin{equation}
    D_{\mu}(q)=aD_{\mu}(-q)+bD_{-\mu-1}(-iq)~.
\end{equation}
Replacing $q$ with $iq$, above equation takes the form
\begin{equation}\label{linearcomb}
    D_{\mu}(iq)=aD_{\mu}(-iq)+bD_{-\mu-1}(q)~.
\end{equation}
Now using eq.(\ref{parabolicansatz}),(\ref{wexpansion}) and (\ref{linearcomb}), the forms of the two constant coefficients $a$ and $b$ can be found as\footnote{Here we have used the property of Gamma function $\Gamma(-\mu)=\frac{-\sqrt{\pi}~~2^{-\mu-1}\Gamma(\frac{1-\mu}{2})}{\Gamma(1+\frac{\mu}{2})\sin{(\pi\mu/2)}}$.}
\begin{equation}
    a=e^{i\mu\pi}, ~~b=\frac{\sqrt{2\pi}}{\Gamma(-\mu)}e^{i(\mu+1)\pi/2}~.
\end{equation}
In asymptotic limit, the required form of wavefunction would be $\psi\sim e^{-i\mathcal{B}(x-x_{0})^{2}}$ where $\mathcal{B}$ is some function of $Q(x)$ which is a constant in asymptotic limit. Hence in eq.(\ref{linearcomb}), the second term has to be zero and hence $\Gamma(-\mu)=\infty$ that means $\mu$ to be any positive integer $n$. Now using eq.(\ref{functiondef}) and $Q(x)\equiv(\omega^{2}-V(r))$ we get the expression for quasi-normal frequency $\omega$ as
\begin{equation}\label{qnfexp}
    \omega^{2} = V(r_{0})-i(n+\frac{1}{2})\sqrt{-2V^{\prime\prime}(r_{0})}~,~~~~n=0,1,2,...
\end{equation}
where $r_{0}$ represents the extrema of the potential $V(r)$.\\

\subsection{Electromagnetic perturbations}
\begin{center}
\begin{figure}[H]
    \centering
    \begin{subfigure}{0.45\textwidth}
        \centering
        \includegraphics[scale=0.47]{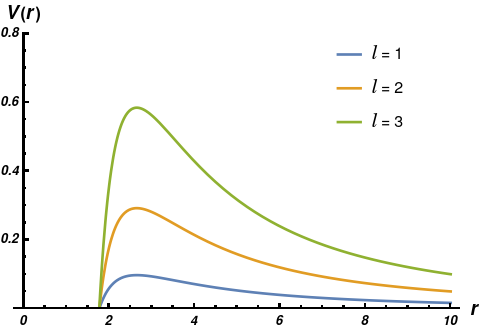}
        \caption{Effective potential for $l=1,2,3$ ($\zeta = 0.07$)}
        \label{plot3}
    \end{subfigure}
    \hfill
    \begin{subfigure}{0.45\textwidth}
        \centering
        \includegraphics[scale=0.47]{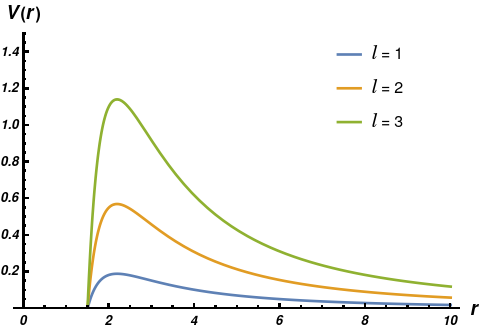}
        \caption{Effective potential for $l=1,2,3$ ($\zeta = 0.80$)}
        \label{plot4}
    \end{subfigure}
    \caption{Comparison of effective potentials for electromagnetic field perturbations for $l = 1, 2, 3$ at different values of $\zeta$.}
    \label{combined_potential_em}
\end{figure}
\end{center}

\noindent In this subsection, we shall discuss the electromagnetic perturbation in spherically symmetric black hole spacetime (eq.(\ref{metric})). The EM field in curved spacetime is governed by the Maxwell's equations
\begin{equation}
    \nabla_{\mu}F^{\mu\nu}=0
\end{equation}
where $F^{\mu\nu}$ is the EM field tensor which is defined by 
\begin{equation}
    F_{\mu\nu}=\partial_{\mu}A_{\nu}-\partial_{\nu}A_{\mu}
\end{equation}
with $A_{\mu}$ being the electromagnetic potential. Hence the wave equation for EM field reads \cite{jusufi}
\begin{equation}\label{emwaveeq}
    \frac{1}{\sqrt{-g}}\partial_{\nu}\left(\sqrt{-g}~g^{\alpha\mu}g^{\sigma\nu}(\partial_{\alpha}A_{\sigma}-\partial_{\sigma}A_{\alpha}) \right)=0~.
\end{equation}
$A_{\mu}$ in spherically symmetric black hole background can be expanded in four-dimensional vector spherical harmonics as \cite{cardosolemos,molinapavan,toshmatov}
\begin{align}\label{empotentialansatz}
    A_{\mu}(t,r,\theta,\phi)=\sum \limits_{l,m} & \left( 
    \begin{bmatrix}
        0 \\
        0\\
        \frac{a_{l,m}(t,r)}{\sin{\theta}}\partial_{\phi}Y_{l,m}(\theta,\phi)\\
        -a_{l,m}(t,r)\sin{\theta}\partial_{\theta}Y_{l,m}(\theta,\phi)
    \end{bmatrix}
    \right. \nonumber \\ &
   \left.  +
    \begin{bmatrix}
        g_{l,m}(t,r)Y_{l,m}(\theta,\phi)\\
        h_{l,m}(t,r)Y_{l,m}(\theta,\phi)\\
        k_{l,m}(t,r)\partial_{\theta}Y_{l,m}(\theta,\phi)\\
        k_{l,m}(t,r)\partial_{\phi}Y_{l,m}(\theta,\phi)
    \end{bmatrix}
    \right)~.
\end{align}
Here, the axial part (first term) has $(-1)^{l+1}$ parity and the polar part (second term) has $(-1)^{l}$ parity.\\
\noindent Considering axial and polar perturbations individually, we obtain the wave equation from eq.(\ref{emwaveeq}) and (\ref{empotentialansatz}) as
\begin{equation}
    \frac{d^{2}\psi}{dr_{*}^{2}}+\left(\omega^{2}-V(r_{*}) \right)\psi=0
\end{equation}
where $r_{*}$ is the tortoise coordinate as defined earlier and $V(r)$ is the effective potential for EM perturbations which has the form 
\begin{equation}\label{potential_em}
    V(r)=f(r)\frac{l(l+1)}{r^{2}}~.
\end{equation}
\noindent In our case, the effective potential takes the form as
\begin{equation}
    V(r)=\left(1-\frac{2GMr}{(r^{2}+\tilde{\omega}G)}+\frac{\zeta}{r}\ln{\left(\frac{r}{|\zeta|}\right)} \right)\times \frac{l(l+1)}{r^{2}}~. 
\end{equation}
\noindent Fig.(\ref{combined_potential_em}) shows the nature of the effective potential $V(r)$ for electromagnetic perturbations. It is obvious that these are also of same nature as for the scalar perturbation was (Fig.(\ref{combined_potential})). The heights of the potential barriers are slightly less in compared to the scalar field perturbations. 
\onecolumngrid
\twocolumngrid
\vspace{0em}
\noindent The wave functions for axial and polar perturbations have the forms \cite{toshmatov,jusufi}
\begin{align}
    &\psi_{a}(t,r)=a_{l,m}(t,r)~~, \nonumber \\ &
    \psi_{p}(t,r)=\frac{r^{2}}{l(l+1)}\left(\partial_{t}h_{l,m}(t,r)-\partial_{r}g_{l,m}(t,r) \right)~.
\end{align}
Now, following the WKB method mentioned in the earlier subsection, we can write the quasi-normal frequency for EM perturbations in terms of the effective potential same as eq.(\ref{qnfexp}).

\section{the effect of PFDM on shadow radius and qnf  }
\begin{figure}[H]
\begin{center}
\includegraphics[scale=0.45]{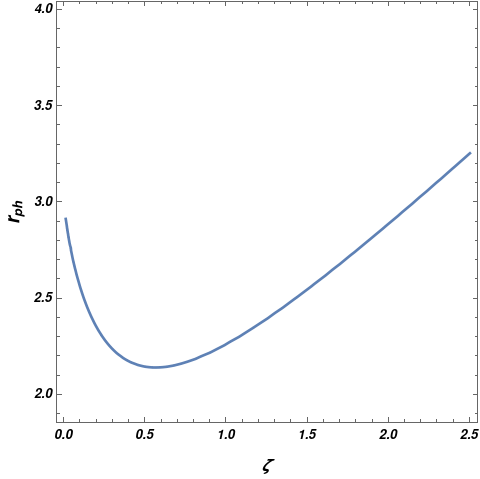}
\caption{Variation of $r_{ph}$ with $\zeta$ \label{plot5}.}
\end{center}
\end{figure}

\noindent In this section, we shall discuss the relation between the quasi-normal frequencies and the shadow radius and how they are affected by the PFDM.\\
\noindent Using the critical orbit conditions (eq.(\ref{criticalorbit})), the equation for photon radius $r_{ph}$ for spherically symmetric black hole is obtained in eq.(\ref{photonradius}).~Considering the lapse function of RG improved Schwarzschild black hole mentioned in eq.(\ref{finalmetric}), we can recast eq.(\ref{photonradius}) as
\begin{equation}\label{rpheqn}
    2r_{ph}^{3}+\left[\zeta\left \lbrace 3\ln{\left(\frac{r_{ph}}{|\zeta|} \right)-1} \right \rbrace-6GM \right]r_{ph}^{2}+10G^{2}M\tilde{\omega}=0
\end{equation}
\noindent where we have taken only the linear order of $\tilde{\omega}$.\\
\noindent Being a transcendental equation, it is difficult to solve the above equation analytically to find the exact expression of $r_{ph}$. In Fig.(\ref{plot5}), we have plotted the photon radius $r_{ph}$ against the PFDM parameter $\zeta$, setting $\tilde{\omega}=0.01$. Initially $r_{ph}$ decreases with increasing $\zeta$ and after reaching a minima, it starts increasing.
\noindent The shadow of the black hole is formed due to the deflection of light it its vicinity. The expression for the shadow radius follows from eq.(\ref{shadowradius}).\\
\begin{table}[H]
\centering
\renewcommand{\arraystretch}{1.1} 
\setlength{\tabcolsep}{14pt}      
\begin{tabular}{c|c}
\hline
$\zeta$ & $R_S$ \\
\hline
0.1 & 4.34832 \\
0.2 & 3.91577 \\
0.5 & 3.35368 \\
0.6 & 3.28651 \\
0.8 & 3.24119 \\
1.0 & 3.27035 \\
1.2 & 3.34430 \\
1.4 & 3.44694 \\
1.6 & 3.56889 \\
1.8 & 3.70442 \\
2.0 & 3.84980 \\
\hline
\end{tabular}
\caption{Variation of $R_S$ with respect to $\zeta$.}
\label{tab1}
\end{table}

\noindent For quantum corrected Schwarzschild black hole surrounded by PFDM, the expression for the shadow radius is
\begin{align}
    R_{s}&=\frac{r_{ph}}{\sqrt{1-\frac{2GMr_{ph}}{(r_{ph}^{2}+\tilde{\omega}G)}+\frac{\zeta}{r_{ph}}\ln{\left(\frac{r_{ph}}{|\zeta|}\right)}}}\nonumber \\ &
    \simeq \frac{\sqrt{3}r_{ph}\biggr[1-\frac{12G^{2}M\tilde{\omega}}{r_{ph}^{3}(1+\frac{\zeta}{r_{ph}})} \biggr]}{\sqrt{1+\frac{\zeta}{r_{ph}}}} ~.
\end{align}
\noindent In the final expression, the logarithmic term $\ln{\left(\frac{r_{ph}}{|\zeta|}\right)}$ has been replaced using eq.(\ref{rpheqn}), keeping contributions only up to first order in $\tilde{\omega}$.

\begin{center}
\begin{figure}[H]
    \centering
    \begin{subfigure}{0.45\textwidth}
        \centering
        \includegraphics[scale=0.47]{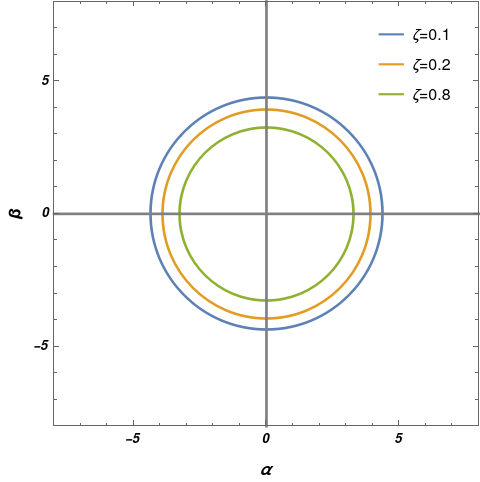}
        \caption{Black hole shadow for different $\zeta\le0.8$}
        \label{plot6}
    \end{subfigure}
    \hfill
    \begin{subfigure}{0.45\textwidth}
        \centering
        \includegraphics[scale=0.47]{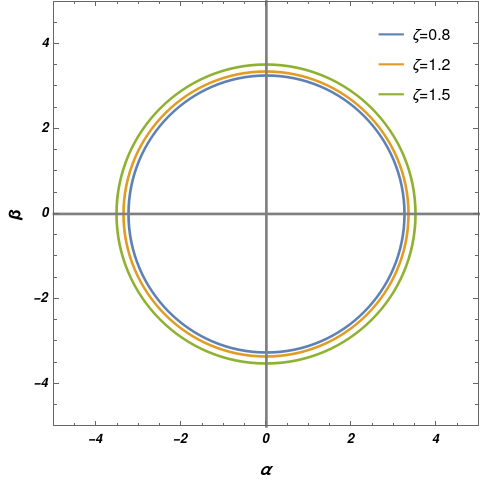}
        \caption{Black hole shadow for different $\zeta \ge 0.8$}
        \label{plot7}
    \end{subfigure}
    \caption{Effect of PFDM parameter on the black hole shadow}
    \label{shadowplot}
\end{figure}
\end{center}

\noindent In Fig.(\ref{shadowplot}), we have shown the effect of PFDM parameter $\zeta$ on the radius of black hole shadow. It is clear from the plot and Table \ref{tab1} that at $\zeta=0.8$, the size of the shadow is minimum, after which the shadow radius starts increasing with $\zeta$.
\onecolumngrid
\twocolumngrid

\begin{center}
\begin{figure}[H]
    \centering
    \begin{subfigure}{0.45\textwidth}
        \centering
        \includegraphics[scale=0.47]{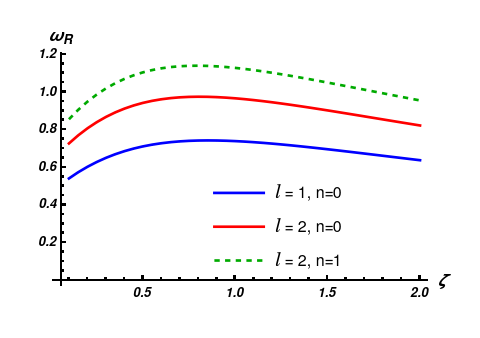}
        \caption{Variation of $\omega_{R}$ for scalar field perturbation with $\zeta$}
        \label{plot8}
    \end{subfigure}
    \hfill
    \begin{subfigure}{0.45\textwidth}
        \centering
        \includegraphics[scale=0.47]{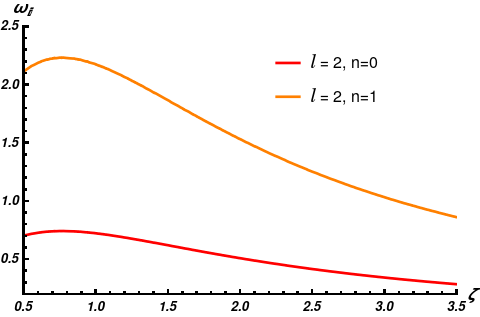}
        \caption{Variation of $\omega_{I}$ for scalar field perturbation with $\zeta$}
        \label{plot9}
    \end{subfigure}
    \caption{Effect of PFDM parameter on the QNF of scalar field perturbation}
    \label{omega_scalar}
\end{figure}
\end{center}

\noindent Now we shall discuss the relation between this photon radius and quasi-normal frequency.\\
\noindent Since, on the photon sphere ($r=r_{ph}$), effective potential has the maxima,  we can write eq.(\ref{qnfexp}) as
\begin{equation}\label{qnfexp2}
    \omega^{2} = V(r_{ph})-i(n+\frac{1}{2})\sqrt{-2V^{\prime\prime}(r_{ph})}\equiv A-iB
\end{equation}
where $A=V(r_{ph})$ and $B=(n+\frac{1}{2})\sqrt{-2V^{\prime\prime}(r_{ph})}$.\\
\noindent Now we can consider the expression for QNF as
\begin{equation}\label{omega}
    \omega = \omega_{R}-i\omega_{I}
\end{equation}
with $\omega_{R}$ and $\omega_{I}$ being the real part and the imaginary part of the QNF respectively.\\
\noindent Taking the square of eq.(\ref{omega}) and comparing with eq.(\ref{qnfexp2}), expressions for $\omega_{R}$ and $\omega_{I}$ can be obtained as
\begin{equation}\label{qnfcomponents}
    \omega_{R}=\sqrt{\frac{A+\sqrt{A^{2}+B^{2}}}{2}}~, ~~\omega_{I}=\frac{B}{\sqrt{2\left(A+\sqrt{A^{2}+B^{2}}\right)}}~.
\end{equation}
\noindent Using the expression of $A$ and $B$ mentioned in eq.(\ref{qnfexp2}), we obtain the forms of real and imaginary parts of quasi-normal frequency in terms of the effective potential as
\begin{align}
    &\omega_{R}=\sqrt{\frac{V(r_{ph})+\sqrt{V^{2}(r_{ph})-2(n+\frac{1}{2})V^{\prime\prime}(r_{ph})}}{2}},\nonumber \\ &
    \omega_{I}=\frac{(n+\frac{1}{2})\sqrt{-2V^{\prime\prime}(r_{ph})}}{2\left(V(r_{ph})+\sqrt{V^{2}(r_{ph})-2(n+\frac{1}{2})V^{\prime\prime}(r_{ph})}\right)}~.
\end{align}
\noindent In Fig.(\ref{omega_scalar}), we have plotted $\omega_{R}$ and $\omega_{I}$ against the PFDM parameter $\zeta$ for different values of $l$ and $n$. It is shown that both $\omega_{R}$ and $\omega_{I}$ has maxima and then decreases with increasing $\zeta$. We plotted $\omega_{R}$ for different values of $l$ and $n$. For large $l$, the peak is higher and between the plots for $l=2$, large $n$ gives the higher peak. In the plot of $\omega_{I}$ (Fig.(\ref{plot9})), higher $n$ value shows the higher peak of quasi-normal frequency.
\begin{table}[H]
\centering
\renewcommand{\arraystretch}{1.1} 
\setlength{\tabcolsep}{14pt}      
\begin{tabular}{c|c|c}
\hline
$\zeta$ & $\omega (l=2, n=0)$ & $\omega (l=2, n=1)$\\
\hline
0.5 & $0.9422-0.7053i$ & $1.1043-2.1159i$\\
0.6 & $0.9614-0.7309i$ & $1.1258-2.1927i$\\
0.8 & $0.9750-0.7438i$ & $1.1397-2.2315i$\\
1.0 & $0.9665-0.7250i$ & $1.1281-2.1751i$\\

1.2 & $0.9454-0.6888i$ & $1.1021-2.0666i$\\

1.4 & $0.9175-0.6444i$ & $1.0684-1.9333i$\\
1.5 & $0.9022-0.6226i$ & $1.0501-1.8680i$\\
1.6 & $0.8864-0.5992i$ & $1.0313-1.7977i$\\

1.8 & $0.8542-0.5537i$ & $0.9931-1.6611i$\\

2.0 & $0.8222-0.5114i$ & $0.9551-1.5342i$\\
\hline
\end{tabular}
\caption{Variation of $\omega$ with respect to $\zeta$ for scalar field perturbation.}
\label{tab2}
\end{table}
\onecolumngrid
\twocolumngrid

\begin{center}
\begin{figure}[H]
    \centering
    \begin{subfigure}{0.45\textwidth}
        \centering
        \includegraphics[scale=0.5]{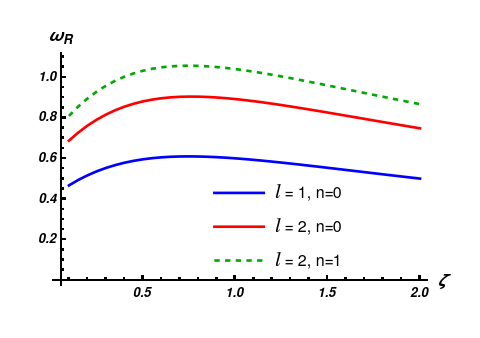}
        \caption{variation of $\omega_{R}$ for electromagnetic field perturbation with $\zeta$}
        \label{plot10}
    \end{subfigure}
    \hfill
    \begin{subfigure}{0.45\textwidth}
        \centering
        \includegraphics[scale=0.47]{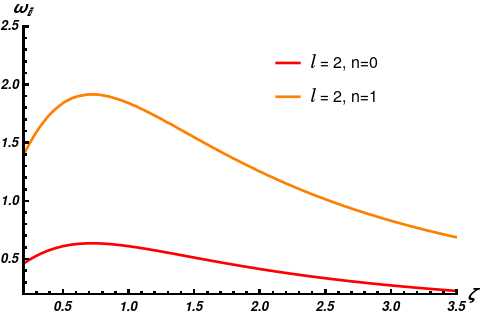}
        \caption{variation of $\omega_{I}$ for electromagnetic field perturbation with $\zeta$}
        \label{plot11}
    \end{subfigure}
    \caption{Effect of PFDM parameter on the QNF of electromagnetic field perturbation}
    \label{omega_em}
\end{figure}
\end{center}

\begin{table}[H]
\centering
\renewcommand{\arraystretch}{1.1} 
\setlength{\tabcolsep}{14pt}      
\begin{tabular}{c|c|c}
\hline
$\zeta$ & $\omega (l=2, n=0)$ & $\omega (l=2, n=1)$\\
\hline
0.5 & $0.8815-0.6158i$ & $1.0325-1.8476i$\\
0.6 & $0.8969-0.6335i$ & $1.0494-1.9005i$\\
0.8 & $0.9049-0.6375i$ & $1.0566-1.9127$\\
1.0 & $0.8930-0.6147i$ & $1.0409-1.8442i$ \\

1.2 & $0.8702-0.5783i$ & $1.0127-1.7349i$\\

1.4 & $0.8418-0.5373i$ & $0.9784-1.6121i$\\
1.5 & $0.8266-0.5165i$ & $0.9601-1.5497i$ \\
1.6 & $0.8110-0.4960i$ & $0.9416-1.4881i$\\

1.8 & $0.7797-0.4553i$ & $0.9043-1.3660i$ \\

2.0 & $0.7488-0.4181i$ & $0.8679-1.2545i$ \\
\hline
\end{tabular}
\caption{Variation of $\omega$ with respect to $\zeta$ for electromagnetic field perturbation.}
\label{tab3}
\end{table}
\noindent Fig.(\ref{omega_em}) shows the variations of quasi-normal frequencies for electromagnetic field perturbations. We have plotted these using eq.(\ref{potential_em}),(\ref{qnfexp2}) and (\ref{qnfcomponents}).  In this case also, both $\omega_{R}$ and $\omega_{I}$ has certain maxima and then they starts decreasing with increasing PFDM parameter $\zeta$. Unlike the previous case, for large $l$ and $n$ the peaks of the quasi-normal frequencies are higher.\\
\noindent Table \ref{tab2} and Table \ref{tab3} present the variation of quasi-normal mode frequency for scalar field and electromagnetic field perturbation respectively. In contrast to the behavior of the shadow radius, the quasi-normal frequencies also exhibit a transition at $\zeta = 0.8$. Both the real part $\omega_R$ and the imaginary part $\omega_I$ increase initially, attain their maximum values at $\zeta = 0.8$, and subsequently decrease beyond this point.

\onecolumngrid
\twocolumngrid

\noindent Comparing two plots (Fig.(\ref{omega_scalar}) and Fig.(\ref{omega_em})), we can see a distinct difference between the two field perturbations. For both the real and imaginary part of the quasi-normal frequency, the peaks are noticeably higher in the case of scalar field perturbation than in the electromagnetic one.\\
\noindent To find the direct relation between quasi-normal frequency and the shadow radius, we shall now impose the eikonal approximation ($l\gg 1$)\cite{cardosomiranda} and the effective potential (eq.(\ref{potential})) can be approximated as
\begin{equation}\label{eikonalpotential}
    V(r)\approx f(r)\frac{l(l+1)}{r^{2}}~.
\end{equation}
\noindent For the lapse function mentioned in eq.(\ref{finalmetric}), the effective potential takes the form in eikonal approximation as
\begin{equation}
    V(r)\approx \left(1-\frac{2GMr}{(r^{2}+\tilde{\omega}G)}+\frac{\zeta}{r}\ln{\left(\frac{r}{|\zeta|}\right)} \right)\times \frac{l(l+1)}{r^{2}}~. 
\end{equation}

\onecolumngrid
\begin{center}
\begin{figure}[H]
    \centering
    \begin{subfigure}{0.45\textwidth}
        \centering
        \includegraphics[scale=0.49]{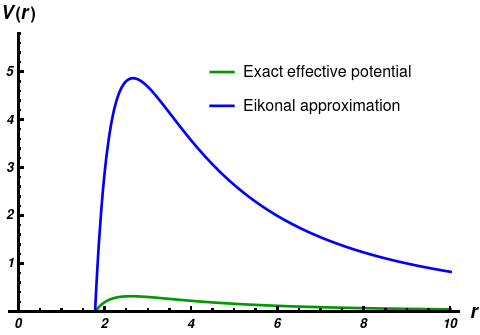}
        \caption{ Scalar field perturbation }
        \label{plot12}
    \end{subfigure}
    \hfill
    \begin{subfigure}{0.45\textwidth}
        \centering
        \includegraphics[scale=0.47]{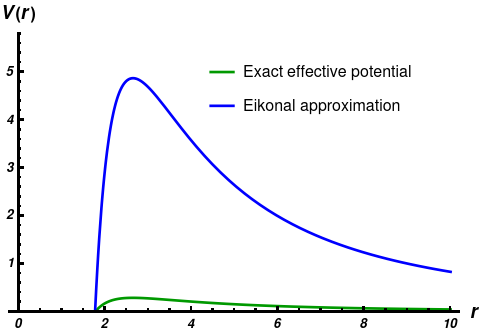}
        \caption{Electromagnetic field perturbation}
        \label{plot13}
    \end{subfigure}
    \caption{Comparing the effective potential in exact form and in eikonal approximation. The exact effective potential has been plotted for $l=2$ and for eikonal approximation, the plot is for $l=10$.}
    \label{eikonal}
\end{figure}
\end{center}
\twocolumngrid
\noindent In Fig.(\ref{eikonal}), we compare the effective potential $V(r)$ obtained from its exact expression with that derived under the eikonal approximation. In the eikonal limit, a larger value of $l$ is considered, resulting in a higher potential peak compared to the exact effective potential, which is plotted for $l=2$. 
\noindent Since in eikonal approximation, the effective potential for both the perturbation has the same form in eq.(\ref{eikonalpotential}), it results eq.(\ref{qnfexp2}) as
\begin{equation}\label{qnfexp3}
    \omega^{2}=l(l+1)\Omega_{ph}^{2}-2i(n+\frac{1}{2})\lambda \Omega_{ph}\sqrt{l(l+1)}
\end{equation}
where $\Omega_{ph}$ is the orbital velocity of the massless scalar photon, defined by
\begin{equation}\label{orbitalvel}
    \Omega_{ph}=\frac{d\phi}{dt}\biggr|_{r=r_{ph}}=\sqrt{\frac{f(r_{ph})}{r_{ph}^{2}}}~.
\end{equation}
$\lambda$ in eq.(\ref{qnfexp3}) is the Lyapunov exponent\footnote{It is different from the affine parameter $\lambda$ mentioned earlier.} which is related to the second order derivative of effective potential on the photon sphere \cite{cardosomiranda}
\begin{equation}
    \lambda=\sqrt{\frac{V^{\prime\prime}(r)}{2\dot{t}^{2}}}\biggr|_{r=r_{ph}}~.
\end{equation}
\noindent Now in large $l$ limit, using eq.(\ref{qnfcomponents}) and (\ref{qnfexp3}), a simplified form of $\omega$ can be written as 
\begin{equation}
    \omega=l\Omega_{ph}-i(n+\frac{1}{2})|\lambda|~.
\end{equation}
\onecolumngrid
\begin{center}
\begin{figure}[H]
    \centering
    \begin{subfigure}{0.45\textwidth}
        \centering
        \includegraphics[scale=0.49]{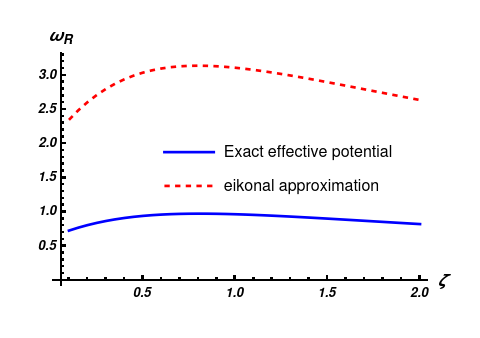}
        \caption{ Scalar field perturbation }
        \label{plot14}
    \end{subfigure}
    \hfill
    \begin{subfigure}{0.45\textwidth}
        \centering
        \includegraphics[scale=0.47]{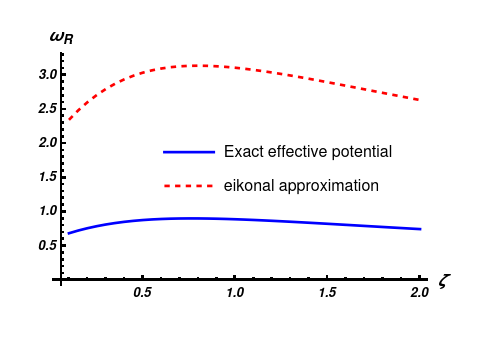}
        \caption{Electromagnetic field perturbation}
        \label{plot15}
    \end{subfigure}
    \caption{Comparing the real part of QNF for exact $V(r)$ and $V(r)$ in eikonal approximation. The exact effective potential has been plotted for $l=2$ and for eikonal approximation, the plot is for $l=10$.}
    \label{qnfeikonal}
\end{figure}
\end{center}
\twocolumngrid
\noindent Fig.(\ref{qnfeikonal}) illustrates the quasi-normal frequencies obtained from the exact effective potential $V(r)$ alongside those derived using the eikonal approximation. In the eikonal regime, the peaks of the QNF are noticeably higher compared to those corresponding to the exact potential.\\
\noindent Now using eq.(\ref{shadowradius}) and eq.(\ref{orbitalvel}), the above expression can be written in terms of shadow radius as
\begin{equation}
    \omega=\frac{l}{R_{s}}-i(n+\frac{1}{2})|\lambda|~.
\end{equation}
\noindent Hence, in eikonal regime, the real part of QNF is inversely proportional to the shadow radius. The importance of this relation relies on the fact that one can measure both the quantities from the knowledge of the other. 
\section{conclusion}
\noindent In this paper, we have studied the effects of field perturbations in the renormalization group improved Schwarzschild black hole background surrounded by perfectly fluid dark matter (PFDM). We have also discussed the effects of PFDM on the shadow of the black hole.\\
\noindent In section II, we have started with the study of the null geodesics of a massless scalar photon in a static, spherically symmetric geometry and using the conditions of critical orbits, we have written down the equation for the radius of the photon sphere in this geometry. In the following section, we have revisited the discussion on theoretical basis for the construction of black hole shadow and the relation between the shadow radius $R_{s}$ and the radius of the photon sphere $r_{ph}$. In section IV, the metric for the Schwarzschild black hole in the presence of PFDM has been derived and the renormalization group improvement of this metric has been discussed. Our present analysis is based on the metric mentioned in eq.(\ref{finalmetric}), which represents the quantum gravity improved Schwarzschild black hole surrounded by PFDM. Here, the Newton's gravitational constant flows with the momentum cut-off scale, which is originated from the renormalization group flow equation. \\
\noindent In section V, we have two subsections. In the first part, we have considered the scalar field perturbations in the RG improved Schwarzschild black hole spacetime in the presence of PFDM. Using the effective potential in eq.(\ref{potential}), we have showed the nature of $V(r)$ for different multipole number $l$, and for larger $l$, the peak of the potential barrier is higher. We have compared the nature of effective potential for $\zeta=0.07$ and $\zeta=0.8$ and it is clearly shown in Fig.(\ref{combined_potential}), that for large $\zeta$, the potential has higher peak. \\
\noindent In the second part, we have considered the electromagnetic field perturbations in the same background. In this case, the effective potential has the form given 
in eq.(\ref{potential_em}). As we have done for the scalar field perturbation, here also we have showed the nature of $V(r)$ for $l=1,2,3$. From the plot (Fig.(\ref{combined_potential_em})), it is clear that for large $l$, the peak of the potential barrier is higher. In this case also, larger value of $\zeta$ gives the higher peaks of the potential barrier.\\
\noindent In the last section, we have studied the effects of perfectly fluid dark matter on the photon radius and the shadow radius in Fig.(\ref{plot5}) and Fig.(\ref{shadowplot}) respectively. The photon radius $r_{ph}$ starts decreasing initially and after reaching a minima, it starts increasing with PFDM parameter $\zeta$. The turning point for $r_{ph}$ is at $\zeta\approx 0.8$, from where it starts to increase. \\
\noindent We have plotted the shadow radius for two cases. First case is for $\zeta\le 0.8$ , while the second one is for $\zeta\ge 0.8$. From the plot, it is clear that for $\zeta=0.8$, the size of the shadow is minimum. The reason for such an observation is that $R_{s}$ is related to $r_{ph}$ via eq.(\ref{shadowradius}).
In this section, we have also showed the effect of $\zeta$ on the real and imaginary parts of the quasi-normal frequencies. We have showed the variations of $\omega_{R}$ and $\omega_{I}$ for both the perturbations with respect to $\zeta$ for different values of $l$ and $n$. $\omega_{R}$ has the higher peak for large $l$ and for $l=2,n=0$ and $l=2,n=1$ plots, large $n$ shows the higher peak. Similar nature is observed in the case of imaginary part of the frequency. There is a distinct difference between the plots for scalar and electromagnetic perturbations. For the electromagnetic field, the frequency peaks are visibly lower than the case for the scalar field. In Fig.(\ref{eikonal}), we compare the exact effective potential with its eikonal approximation. Since the eikonal limit corresponds to large $l$, the peak of the effective potential in this regime is higher than that of the exact potential. A similar analysis is performed for the real part of the QNF in Fig.(\ref{qnfeikonal}), where we again observe that the transition point in the eikonal limit occurs at a higher value compared to the exact case.

\end{document}